\documentclass[prl,twocolumn,showpacs]{revtex4}
\usepackage{graphicx}

\begin{document}

\preprint{}

\title{Self healing slip pulses along a gel/glass interface}

\author{Tristan Baumberger}
\author{Christiane Caroli}
\author{Olivier Ronsin}
 \email{ronsin@gps.jussieu.fr}
\affiliation{Groupe de Physique des Solides, Universit\'es Paris 6 et 7, UMR CNRS 7588, 2 place Jussieu, 75251 Paris, Cedex 05, France}

\date{\today}

\begin{abstract}
We present an experimental evidence of self-healing shear cracks at a gel$/$glass interface. This system exhibits two dynamical regimes depending on the driving velocity~: steady sliding at high velocity ($> V_{\text{c}} \simeq 100$---$125~\mu$m$/$s), caracterized by a shear-thinning rheology, and periodic stick-slip dynamics at low velocity. In this last regime, slip occurs by propagation of pulses that restick via a ``healing instability'' occuring when the local sliding velocity reaches the macroscopic transition velocity $V_{\text{c}}$. At driving velocities close below $V_{\text{c}}$, the system exhibits complex spatio-temporal behavior.
\end{abstract}

\pacs{62.20.Mk, 62.20.Qp, 46.50.+a, 46.55.+d}

\maketitle

There has been in the last decade a sustained interest in the question of the existence, characteristics and dynamical behavior of slip pulses propagating along sheared frictional interfaces. The relevance of such ``self-healing cracks'' to seismic dynamics was first demonstrated by Heaton~\cite{heaton} from inversion of seismic data. They have been evidenced as playing an important role in the complex sliding dynamical behavior of Burridge-Knopoff types of models~\cite{langer}, and are being studied by mechanicians~\cite{rice}, who have recently proved the essential importance, in this context, of the detailed features of the underlying friction law. These multifarious developments are to be contrasted with the much less advanced state of experimental studies in this field. Rubio and Galeano~\cite{rubio}, studying gels sliding against plexiglass (lucite) in a Couette geometry, observed a regime of inhomogeneous sliding via such self-healing pulses. These have also been found by Anooshehpoor and Brune~\cite{brune} with an interface between dissimilar foam rubbers. However, no systematic charaterization of the range of driving velocities where they exist, nor of their detailed structure is yet, to our knowledge, available.

In this Letter, we report the results of such a study, perfomed on a gel (aqueous gelatin) sliding on glass, in a linear geometry. Total sliding force measurements evidence at low driving velocity $V$, a regime of periodic stick-slip which bifurcates, at $V = V_{\text{c}}\simeq 100\ \mu$m/s, towards stationary sliding. Simultaneous optical observation allows us to show that, in the stick-slip regime, slip is {\it inhomogeneous}. It occurs via the propagation of self healing pulses with no observable opening (in contrast with Schallamach waves~\cite{schallamach}), nucleated periodically at the trailing edge of the sliding block. The measurement of the sliding velocity profile within these pulses shows that resticking is determined by a ``healing instability'' occurring at the above mentioned critical velocity $V_{\text{c}}$. This points towards the intimate connection between pulse existence and the detailed form of the velocity dependence of the dynamic frictional stress. Moreover, the velocity of the fracture-like pulse front can be related with small scale characteristics of the gel, namely its mesh size --- in agreement with qualitative arguments put forward by Tanaka \textit{et al.}~\cite{sekimoto} to account for mode I fracture dynamics in a bulk gel. Finally, in a narrow velocity bracket about $V_{\text{c}}$, {\it irregular stick-slip} is observed. It corresponds to the alternation of pulses of the type described above and of shorter lived events nucleated ``homogeneously'' within the interface.

\begin{figure}
\includegraphics{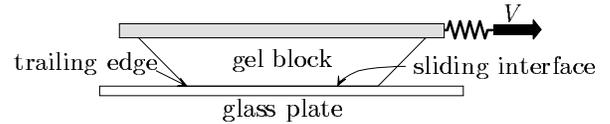}
\caption{\label{fig1}Sketch of the experimental setup.}
\end{figure}

{\it Experiments---}
The gel samples are made from a 5~wt~\% solution of gelatin in water, stirred for 35 minutes at 55$^{\circ}$~C, then poured into a parallelipedic mould of thickness $10$~mm, and kept at 5$^{\circ}$~C for 20 hours. After returning to room temperature, the block sides are cut to obtain the trapezoidal shape shown on figure~\ref{fig1}. This ensures that sliding will occur at the gel$/$glass interface rather than along the upper driving plate. All experiments are performed at $20^{\circ}$ C in a water saturated atmosphere. No measurable ageing of the samples occurs over the duration ($\simeq 3$ hours) of an experiment.
We measure for the low frequency shear modulus $G = 4$~kPa. Moreover, we determined from dynamic light scattering the collective diffusion coefficient of the gelatin network~\cite{tanaka} $D = 1.8\ 10^{-11}$~m$^{2}$.s$^{-1}$. 

The flat free surface of the sample (area $A = 30\times 10$ mm$^{2}$) is brought to contact with a carefully cleaned float glass plate. Fine control of the parallelism of the approaching surfaces, together with  the strong gel$/$glass adhesion and the large compliance of the gel block, ensure the homogeneity of the contact. 
The system is driven at a controlled velocity $V$ ranging from $10~\mu$m$/$s to $2$~mm$/$s through a double cantilever spring much stiffer than the sample. The shear force $F$, and thus the average shear stress $\sigma = F/A$, is deduced from the spring elongation, measured by a capacitive displacement transducer. In addition, a stiff attachment is used to work at a constant sample thickness corresponding to zero external load on the non-moving system.
All experiments consist of several runs of slid distance $150$~mm. The shear stress level is reproducible to better than 5\%, except for the first run, which we discard since it corresponds to a systematically larger stress. 
The interfacial plane is observed with a CCD camera at a rate of 60 frames per second. Its motion can be observed thanks to the presence of tiny optical imperfections attached to the gel in the vicinity of its surface. This permits to measure interfacial slip with a resolution of $20~\mu$m. Finally, sideways optical observation between crossed polarizers provides information about the deformation field intergrated across the sample.\\

\begin{figure}
\includegraphics[draft=false]{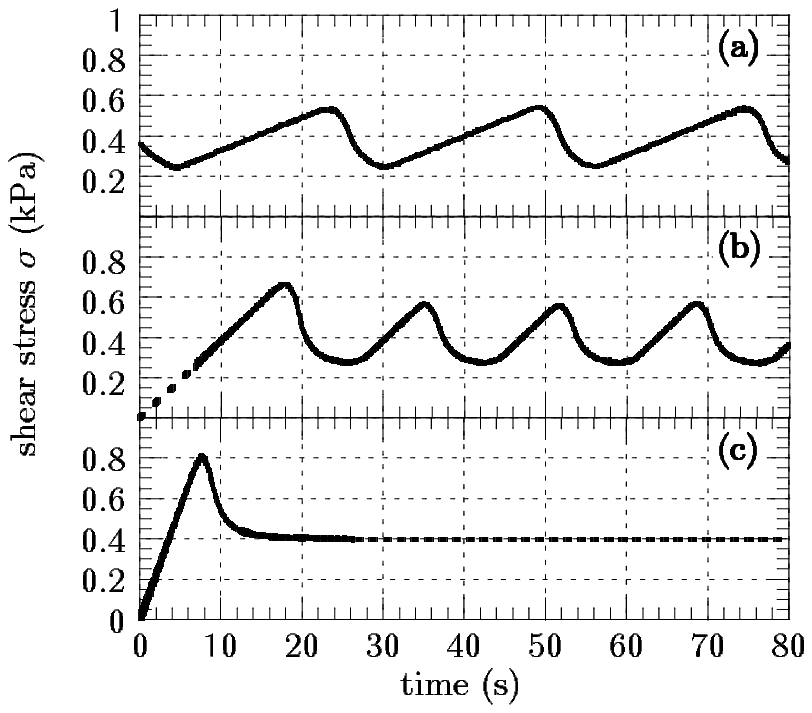}
\caption{\label{fig2}Average shear stress $\sigma = F/A$ {\it vs.} time showing the two dynamical behaviors of the system : Stick-slip at low driving velocity (a) : $V = 50~\mu$m$/$s, (b) $V = 100~\mu$m$/$s. (c) Steady sliding at high velocity $V = 150~\mu$m$/$s.}
\end{figure}

\begin{figure}
\includegraphics{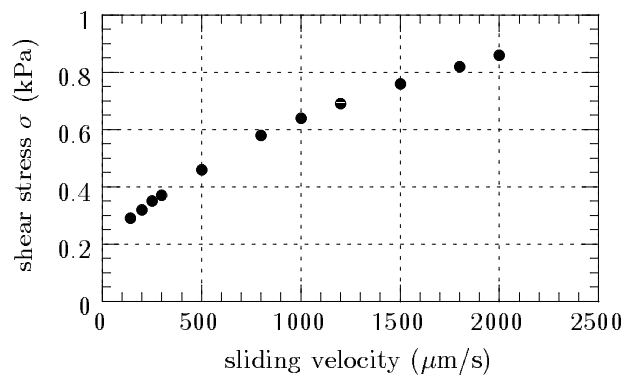}
\caption{\label{fig3}Average shear stress {\it vs.} the sliding velocity in the steady sliding regime ($V > V_{\text{c}}$).}
\end{figure}

When starting to load from rest at constant $V$, one first observes a linear rise of the average shear stress $\sigma$, corresponding to elastic deformation of the block. This stick phase is followed by a slip one, during which $\sigma$ drops. Following this transient stress peak, two dynamical regimes are observed, depending on $V$~: {\it (i)}~ at low velocities, slip stops, and periodic stick-slip sets up (Fig~\ref{fig2}.a). As $V$ is increased, the stress tail during slip becomes longer (Fig~\ref{fig2}.b). The first peak is systematically larger than the subsequent ones, indicating interfacial ageing at rest~\cite{ronsin}. {\it (ii)}~For $V > V_{\text{c}}\simeq 125~\mu$m$/$s, no resticking occurs any more, sliding becomes stationary (Fig~\ref{fig2}.c). In this regime, $\sigma(V)$ (Fig.\ref{fig3}) is measured to be velocity strengthening, of the shear-thinning type frequently observed for complex fluids.

Upon decreasing $V$, {\it no hysteresis of the transition is observed}. However, in a narrow velocity range below $V_{\text{c}}$, the above described stick-slip peaks are aperiodic and interspeded with irregular smaller events (see below). Such a mechanical behaviour seems incompatible with a one-degree of freedom dynamics. This points toward the need for improving upon the integrated information provided by $\sigma$. This is realized by extracting the spatio-temporal slip dynamics from the CCD recording of the interface image. 

\begin{figure*}
\includegraphics[draft=false]{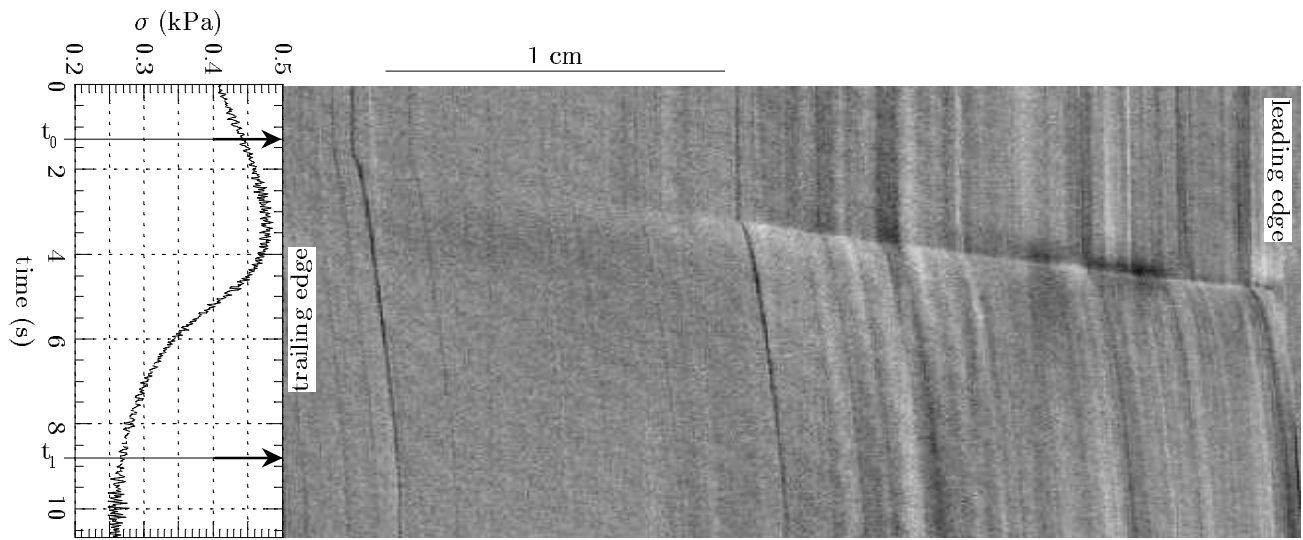}
\caption{\label{fig4}Spatio-temporal diagram of interfacial motion in the stick-slip regime, with time going down, and the gel driven to the right. Dark lines correspond to the trajectories of imperfections attached to the gel surface. For $t < t_0$, they are immobile, and the gel is stuck. At $t_0$, the trailing edge begins to slide, and sliding propagates along the interface. $t_1$ marks the beginning of resticking. The curve on the left shows the average shear stress measured simultaneously.}
\end{figure*}

Figure~\ref{fig4} shows the light intensity along a single pixel line parallel to the driving direction (horizontal axis) as a function of time (vertical axis) in the stick-slip regime ($V = 80~\mu$m$/$s). At the end of the stick phase ($t = t_0$) sliding starts from the trailing edge and propagates in the driving direction. The locus of slope discontinuities reveals the propagation of the head of this ``slip wave''. Its velocity $V_{\text{tip}}$ increases, as intuitively expected, as it comes close to the leading edge. During most of the traversal, $V_{\text{tip}} \simeq 8$~mm$/$s. This value is found to be, within experimental precision, independent of the driving velocity.

\begin{figure}
\includegraphics[draft=false]{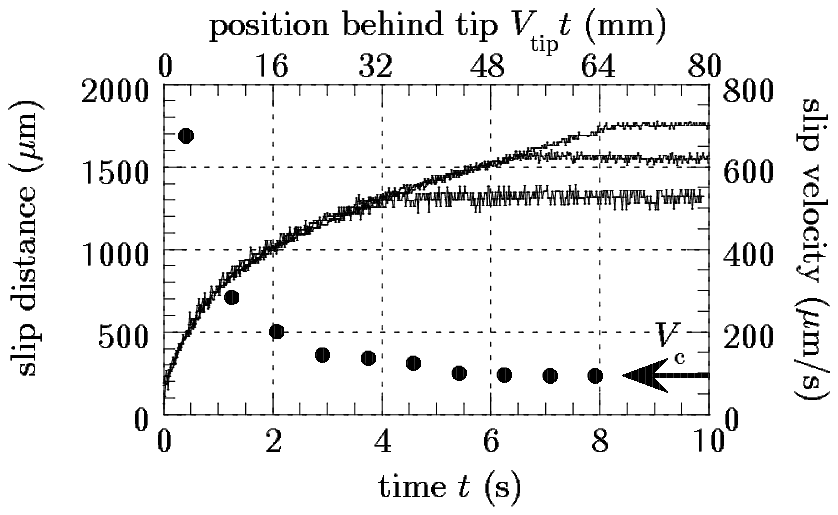}
\caption{\label{fig5}{\it Full curves} : Slip distance {\it vs.} time (lower horizontal axis) and distance behind pulse tip $V_{\text{tip}}t$ (upper horizontal axis). $t = 0$ reffers to the local onset of sliding. From bottom to top $V = 30$, $60$, $90\ \mu$m.s$^{-1}$. {\it Dots} : Velocity profile corresponding to the upper slip curve.}
\end{figure}

We find that all trajectories can be superimposed by shifting into coincidence the points of discontinuity at which sliding starts. That is, the slip field is stationary in the frame of the moving tip. It is shown on Fig.~\ref{fig5}, together with the slip velocity field. Three important qualitative features emerge~: {\it (i)}~the slip velocity exhibits a quasi divergence at the tip, suggestive of a fracture-like singularity. This is confirmed by the photoelastic visualization, which clearly reveals stress concentration in the tip region. It is worth noticing, at this stage, that no opening of the interfacial contact is observable, which leads us to think that we are dealing with a pure ``frictional shear crack'' (see discussion). {\it (ii)}~Surprisingly, resticking occurs quasi-discontinously. Namely, on a time scale $\Delta t < 1/60$~s, the sliding velocity decreases from the finite value $V_{\text{stop}}$ to zero. Moreover, whatever the driving velocity in the stick-slip regime, we find that $V_{\text{stop}}$ {\it is equal to the critical velocity} $V_{\text{c}}$ which marks the transition to the regime of stationary block sliding.{\it(iii)} As seen on Fig.~\ref{fig5}, for $V<V_{\text{c}}$ the slip field close behind the tip is quasi $V$-independent, the slip velocity decreasing asymptotically towards $V_{\text{c}}$. This is reached at a distance which increases with $V$. This increase of the pulse width accounts for the growth of the stick-slip period.

When $V>V_{\text{c}}$, the slip velocity behind the initial pulse head saturates at the driving velocity, and stationary motion sets in after the tip has emerged on the leading edge.

\begin{figure*}
\includegraphics[draft=false]{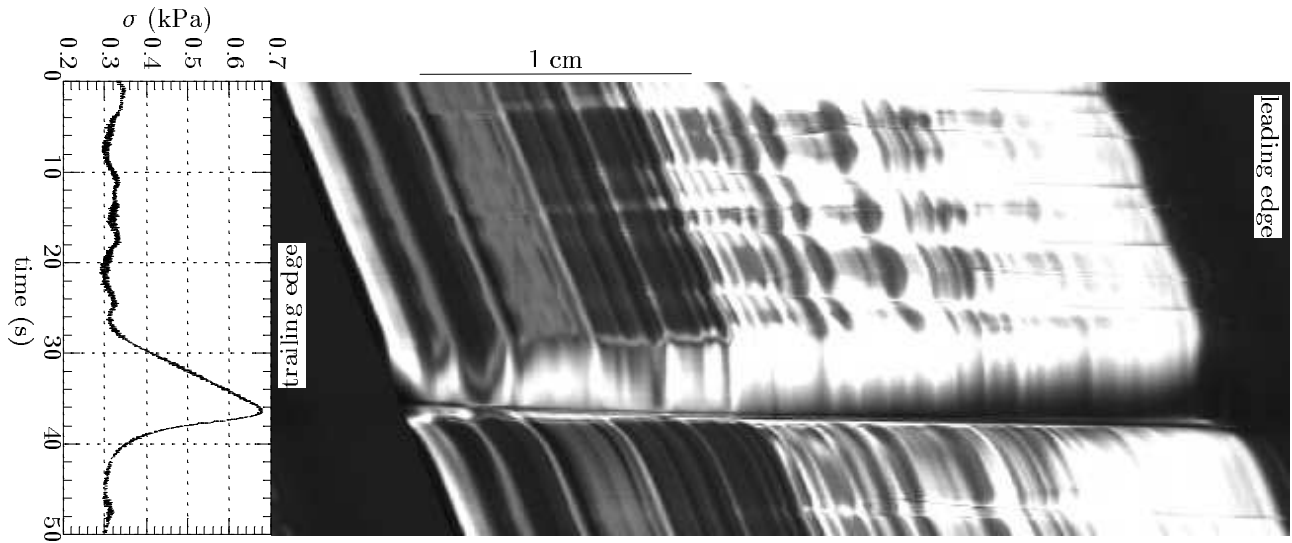}
\caption{\label{fig6}Spatio-temporal diagram at a driving velocity close below $V_{\text{c}}$ built from a line of the photoelastic sideview of the gel block close to the interface. The presence of thickness defects on the lateral faces of the block reveals local sliding, as in figure~\ref{fig3}. Photoelasticity shows waves of deformation of various sizes associated with local slip pulses responsible for the fluctuations of the macroscopic shear stress (curve on the left).}
\end{figure*}

Finally, in a $V$-range of order $5\ \mu$m$/$s below $V_{\text{c}}$, where the pulse width has become larger than the sample one, a more complex behavior is observed (see Fig~\ref{fig6}). Stationary sliding seemingly sets in. However, after a random time lapse, the interface resticks suddenly and practically all at once (no resticking front is observable). The average stress then starts rising again until a new pulse is nucleated at the trailing edge. Moreover, during the sliding phases between these large pulses, finite interfacial patches restick at random. Then, within these patches, slip nucleates and propagates in either direction. These ``homogeneously nucleated'' events are associated with the stress fluctuations interspeding the main peaks.

{\it Discussion---} So, it appears clearly that low-velocity sliding of the interface between our two elastically very dissimilar materials occurs via the ``heterogeneous nucleation'' of self-healing pulses. These sweep the contact at a quasi $V$-independent velocity $V_{\text{tip}}$ on the order of 8~mm$/$s, much slower than the (transverse) sound velocity $c_{\text{s}}\simeq 2$~m$/$s. As was already noticed by Tanaka {\it et al.}~\cite{sekimoto} when studying propagation of mode I fractures in a gel, this strongly subsonic value of $V_{\text{tip}}$ points towards the fact that the main relevant dissipative mechanism in our poroelastic medium is the so-called ``slow Biot mode'' i.e. collective diffusion of the gelatin network. Indeed, this mode, of dispersion law $\omega_{\text{D}} = Dq^2$, is expected to become most effective when the molecular depinning signal from the crack head of frequency $\omega = V_{\text{tip}}/d$ (with $d$ the gel mesh size), resonates with the diffusive mode at wavevector $q\simeq d^{-1}$, i.e. when $V_{\text{tip}}\simeq D/d$. $D$ is related to the gel shear modulus $G$, skeleton Poisson ratio $\nu$ and mesh size $d$ by~\cite{tanaka}:
\begin{equation}
\label{diffu}
D\simeq 2G\frac{1-\nu}{1-2\nu}\frac{d^2}{\eta}
\end{equation}
with $\eta$ the viscosity of water. One can safely assume~\cite{tanaka} that $\nu = 0$. From our measured values of $D$ and $G$, we then estimate $d\simeq 1.5$~nm, so that $D/d\simeq 10$~mm$/$s, in excellent agreement with the measured $V_{\text{tip}}$.

The fact that we observe, for driving velocities close to $V_{\text{c}}$, sudden global resticking events strongly suggests that $V_{\text{c}}$ is a threshold below which sliding becomes unstable down to wavelengths smaller than our space resolution. This inference is consistent with the quasi discontinuity of the slip velocity at the resticking pulse tail end in the periodic regime. In other words, we believe that, if stationary sliding could be stabilized at $V < V_{\text{c}}$ by e.g. working with a very thin, hence stiff, gel layer, sliding friction should become velocity weakening.

Such a behavior must necessarily be ascribed to the underlying dynamics of a structural ``state'' variable. In the present case, one can reasonably associate it with the density of polymer chains pinned to strong adhesion sites (e.g. silanols) on the glass surface which, according to Charitat and Joanny~\cite{joanny}, should decrease with increasing slip velocity. Note that, in this frame of interpretation, adhesion persists while sliding, and opening of the network$/$glass contact is excluded---in contrast with Schallamach waves in rubber or with ``brittle'' pulses studied by Gerde and Marder~\cite{marder}.

Our results point to the importance, for the dynamics and, possibly, the existence, of self-healing pulses in our system, of the details of the frictional dynamics. Whether or not the relation between self-healing and sliding instability might extend to a wider class of systems of course remains an open question. However, it is worth mentioning, in this perspective, the striking qualitative analogy between the global stick-slip dynamics of our system and that measured in some Surface Force Apparatus experiments~\cite{israelashvili} involving thin confined layers of small molecules.

\begin{acknowledgments}
We are indebted to L.~Legrand for his help with the light scattering determination of $D$. We thank J.-F.~Joanny, J.R.~Rice, J.W.~Rudnicky and K.~Sekimoto for illuminating discussions.
\end{acknowledgments}


\begin{thebibliography}{12}
\expandafter\ifx\csname natexlab\endcsname\relax\def\natexlab#1{#1}\fi
\expandafter\ifx\csname bibnamefont\endcsname\relax
  \def\bibnamefont#1{#1}\fi
\expandafter\ifx\csname bibfnamefont\endcsname\relax
  \def\bibfnamefont#1{#1}\fi
\expandafter\ifx\csname citenamefont\endcsname\relax
  \def\citenamefont#1{#1}\fi
\expandafter\ifx\csname url\endcsname\relax
  \def\url#1{\texttt{#1}}\fi
\expandafter\ifx\csname urlprefix\endcsname\relax\def\urlprefix{URL }\fi
\providecommand{\bibinfo}[2]{#2}
\providecommand{\eprint}[2][]{\url{#2}}

\bibitem[{\citenamefont{Heaton}(1990)}]{heaton}
\bibinfo{author}{\bibfnamefont{T.}~\bibnamefont{Heaton}},
  \bibinfo{journal}{Phys.\ Earth\ Planet.\ Inter.}
  \textbf{\bibinfo{volume}{64}}, \bibinfo{pages}{1} (\bibinfo{year}{1990}).

\bibitem[{\citenamefont{Carlson et~al.}(1994)\citenamefont{Carlson, Langer, and
  Shaw}}]{langer}
\bibinfo{author}{\bibfnamefont{J.}~\bibnamefont{Carlson}},
  \bibinfo{author}{\bibfnamefont{J.}~\bibnamefont{Langer}}, \bibnamefont{and}
  \bibinfo{author}{\bibfnamefont{B.}~\bibnamefont{Shaw}},
  \bibinfo{journal}{Rev.\ Mod.\ Phys.} \textbf{\bibinfo{volume}{66}},
  \bibinfo{pages}{657} (\bibinfo{year}{1994}), \bibinfo{note}{and references
  therein}.

\bibitem[{\citenamefont{Ranjith and Rice}(2001)}]{rice}
\bibinfo{author}{\bibfnamefont{K.}~\bibnamefont{Ranjith}} \bibnamefont{and}
  \bibinfo{author}{\bibfnamefont{J.}~\bibnamefont{Rice}}, \bibinfo{journal}{J.
  Mech. Phys. Solids} \textbf{\bibinfo{volume}{49}}, \bibinfo{pages}{341}
  (\bibinfo{year}{2001}).

\bibitem[{\citenamefont{Rubio and Galeano}(1994)}]{rubio}
\bibinfo{author}{\bibfnamefont{M.}~\bibnamefont{Rubio}} \bibnamefont{and}
  \bibinfo{author}{\bibfnamefont{J.}~\bibnamefont{Galeano}},
  \bibinfo{journal}{Phys. Rev. E} \textbf{\bibinfo{volume}{50}},
  \bibinfo{pages}{1000} (\bibinfo{year}{1994}).

\bibitem[{\citenamefont{Anooshehpoor and Brune}(1994)}]{brune}
\bibinfo{author}{\bibfnamefont{A.}~\bibnamefont{Anooshehpoor}}
  \bibnamefont{and} \bibinfo{author}{\bibfnamefont{J.}~\bibnamefont{Brune}},
  \bibinfo{journal}{PAGEOPH} \textbf{\bibinfo{volume}{143}},
  \bibinfo{pages}{61} (\bibinfo{year}{1994}).

\bibitem[{\citenamefont{Schallamach}(1971)}]{schallamach}
\bibinfo{author}{\bibfnamefont{A.}~\bibnamefont{Schallamach}},
  \bibinfo{journal}{Wear} \textbf{\bibinfo{volume}{17}}, \bibinfo{pages}{301}
  (\bibinfo{year}{1971}).

\bibitem[{\citenamefont{Tanaka et~al.}(1999)\citenamefont{Tanaka, Fukao,
  Miyamoto, and Sekimoto}}]{sekimoto}
\bibinfo{author}{\bibfnamefont{Y.}~\bibnamefont{Tanaka}},
  \bibinfo{author}{\bibfnamefont{K.}~\bibnamefont{Fukao}},
  \bibinfo{author}{\bibfnamefont{Y.}~\bibnamefont{Miyamoto}}, \bibnamefont{and}
  \bibinfo{author}{\bibfnamefont{K.}~\bibnamefont{Sekimoto}},
  \bibinfo{journal}{Europhys. Lett.} \textbf{\bibinfo{volume}{43}},
  \bibinfo{pages}{664} (\bibinfo{year}{1999}).

\bibitem[{\citenamefont{Tanaka et~al.}(1973)\citenamefont{Tanaka, Hocker, and
  Benedek}}]{tanaka}
\bibinfo{author}{\bibfnamefont{T.}~\bibnamefont{Tanaka}},
  \bibinfo{author}{\bibfnamefont{L.}~\bibnamefont{Hocker}}, \bibnamefont{and}
  \bibinfo{author}{\bibfnamefont{G.}~\bibnamefont{Benedek}},
  \bibinfo{journal}{J. Chem. Phys.} \textbf{\bibinfo{volume}{59}},
  \bibinfo{pages}{5151} (\bibinfo{year}{1973}).

\bibitem[{\citenamefont{Baumberger et~al.}(2001)\citenamefont{Baumberger,
  Caroli, and Ronsin}}]{ronsin}
\bibinfo{author}{\bibfnamefont{T.}~\bibnamefont{Baumberger}},
  \bibinfo{author}{\bibfnamefont{C.}~\bibnamefont{Caroli}}, \bibnamefont{and}
  \bibinfo{author}{\bibfnamefont{O.}~\bibnamefont{Ronsin}}
  (\bibinfo{year}{2001}), \bibinfo{note}{(to be published)}.

\bibitem[{\citenamefont{Charitat and Joanny}(2000)}]{joanny}
\bibinfo{author}{\bibfnamefont{T.}~\bibnamefont{Charitat}} \bibnamefont{and}
  \bibinfo{author}{\bibfnamefont{J.-F.} \bibnamefont{Joanny}},
  \bibinfo{journal}{Aur. Phys. J. E} \textbf{\bibinfo{volume}{3}},
  \bibinfo{pages}{369} (\bibinfo{year}{2000}).

\bibitem[{\citenamefont{Gerde and Marder}(2001)}]{marder}
\bibinfo{author}{\bibfnamefont{E.}~\bibnamefont{Gerde}} \bibnamefont{and}
  \bibinfo{author}{\bibfnamefont{M.}~\bibnamefont{Marder}},
  \bibinfo{journal}{Nature} \textbf{\bibinfo{volume}{413}},
  \bibinfo{pages}{285} (\bibinfo{year}{2001}).

\bibitem[{\citenamefont{Yoshizawa and Israelashvili}(1993)}]{israelashvili}
\bibinfo{author}{\bibfnamefont{H.}~\bibnamefont{Yoshizawa}} \bibnamefont{and}
  \bibinfo{author}{\bibfnamefont{J.}~\bibnamefont{Israelashvili}},
  \bibinfo{journal}{J. Phys. Chem.} \textbf{\bibinfo{volume}{97}},
  \bibinfo{pages}{11300} (\bibinfo{year}{1993}).

\end{thebibliography}
\end{document}